\documentclass[11pt,a4paper]{article}

\usepackage[T1]{fontenc}
\usepackage[utf8]{inputenc}
\usepackage{lmodern}
\usepackage[margin=1in]{geometry}
\usepackage{booktabs}
\usepackage{graphicx}
\usepackage{amsmath}
\usepackage{algorithm}
\usepackage{algpseudocode}
\usepackage{microtype}
\usepackage{url}
\usepackage[htt]{hyphenat}  
\usepackage{authblk}
\usepackage[backend=biber,style=numeric,sorting=none,giveninits=true,maxbibnames=6]{biblatex}
\usepackage[hidelinks]{hyperref}

\newcommand{\HdcoDescriptions}{2\,811}
\newcommand{\HdcoCatalogues}{11}
\newcommand{\HdcoObservations}{22\,488}

\newcommand{\HdcoTriples}{221\,431}

\newcommand{\HdcoDateLong}{25 August 2026}
\newcommand{\HdcoChecks}{18}
\newcommand{\HdcoLayerTwo}{21\,431}
\newcommand{\HdcoLayerThree}{11}
\newcommand{\HdcoDefectOne}{12\,975}
\newcommand{\HdcoDefectTwo}{8\,433}
\newcommand{\HdcoFiTotal}{2\,259}
\newcommand{\HdcoFiThemed}{1\,146}

\newcommand{\HdcoIrisInUse}{36}
\newcommand{\HdcoIrisDefined}{14}
\newcommand{\HdcoIrisUndefined}{22}
\newcommand{\HdcoSquatted}{1\,238}
\newcommand{\HdcoUndefinedTerm}{806}
\newcommand{\HdcoEnvTerm}{237}
\newcommand{\HdcoEmptyBytes}{170}
\newcommand{\HdcoHealBytes}{16\,063}
\newcommand{\HdcoFindataDatasets}{2\,835}
\newcommand{\HdcoFindataVariables}{89\,368}
\newcommand{\HdcoFindataNoEnglishVar}{75\,729}
\newcommand{\HdcoFindataNoEnglishPct}{84.74}
\newcommand{\HdcoFindataEnLabel}{213}
\newcommand{\HdcoFindataEnLabelPct}{7.51}
\newcommand{\HdcoFindataNoDescVar}{1\,799}
\newcommand{\HdcoFindataNoDescVarPct}{2.01}
\newcommand{\HdcoFindataConceptTags}{2\,948}
\newcommand{\HdcoFindataConcepts}{508}
\newcommand{\HdcoFindataConceptsEn}{477}
\newcommand{\HdcoSwedenKnown}{2\,418}
\newcommand{\HdcoThemedCats}{10}
\newcommand{\HdcoNoThemeCats}{1}
\newcommand{\HdcoAboveNinety}{6}
\newcommand{\HdcoSecondLowestPct}{32.3}
\newcommand{\HdcoSecondLowestName}{geonorge}
\newcommand{\HdcoMqaMax}{405}
\newcommand{\HdcoMqaBest}{300}

\newcommand{\HdcoMqaExcellentLow}{351}

\newcommand{\HdcoMqaSweScore}{177}
\newcommand{\HdcoMqaSweComp}{27}
\newcommand{\HdcoMqaDkScore}{287}
\newcommand{\HdcoMqaDkComp}{0}
\newcommand{\HdcoMqaZeroComp}{5}
\newcommand{\HdcoMqaRetrieved}{27 August 2026}
\newcommand{\HdcoBasePublisher}{2\,785}

\newcommand{\HdcoLegislationPct}{0.75}

\title{Measuring the Installed Base: Nordic Health Dataset Catalogues\\
Against HealthDCAT-AP Release 7}

\author[1]{Fabio Rovai}
\affil[1]{Kampakis and Co Ltd trading as The Tesseract Academy, London, United Kingdom\\
\texttt{fabio@thetesseractacademy.com}}

\date{27 August 2026}

\begin{document}
\maketitle

\begin{abstract}
\noindent
The European Health Data Space requires member states to publish machine readable
descriptions of the health datasets available for secondary use, and the European
Commission publishes HealthDCAT-AP as the metadata profile those descriptions are
meant to satisfy. The profile has been designed and validated against curated
examples. It has never been measured against the catalogues that are already live.
We report that measurement for the Nordic region. On \HdcoDateLong{} the
\HdcoCatalogues{} Nordic national catalogues harvested by the European data portal
held \HdcoDescriptions{} dataset descriptions carrying the EU health theme, and
none of them satisfies all eight properties HealthDCAT-AP Release 7 makes mandatory
on a dataset. Three of the eight are present on exactly zero records across five
countries. Setting our verdict beside the portal's own quality assessment, which
validates DCAT-AP and does not mention the health profile, shows that this is not a
health extension skipped on top of sound generic practice: no Nordic catalogue
reaches the assessment's top rating band, \HdcoMqaZeroComp{} of the
\HdcoCatalogues{} are reported at zero per cent DCAT-AP compliant, and the
properties that survive in both layers are the ones a human types into a form
rather than the ones requiring a value bound to a controlled vocabulary. Two
further results fall out of the same measurement. Finland contributes
\HdcoFiTotal{} descriptions to the European portal of which \HdcoFiThemed{} carry a
theme, and not one of those uses the EU theme authority vocabulary, so a European
health filter returns no Finnish dataset at all. Separately, the authority
namespace answers HTTP 200 for terms it has never defined and returns a well formed
empty document, which lets \HdcoSquatted{} datasets across the wider portal carry
theme IRIs that resolve to nothing while passing any status code check. We publish
the vocabulary, the shapes and the harvesters, we record every verdict as a dated
observation rather than as a property of the dataset, and we report the five
errors that this discipline caught before publication.
\end{abstract}

\section{Introduction}

The European Health Data Space Regulation entered into force on 26 March
2025~\cite{ehds2025}. Its secondary use provisions oblige member states to make
health data discoverable across borders, and discovery in this architecture runs on
dataset descriptions rather than on the data itself. A researcher in Lisbon does not
query a Finnish register. They query a catalogue of descriptions, decide which
Finnish datasets are worth a permit application, and only then approach the health
data access body. Everything downstream of that first step depends on the
description being present, complete and expressed in vocabulary the querying system
resolves.

The European Commission publishes HealthDCAT-AP for exactly this
purpose~\cite{healthdcatap7}. It is an application profile over DCAT-AP~3.0.1,
itself an application profile over W3C DCAT, and it adds the properties a health
catalogue needs and a generic open data catalogue does not: which health data access
body is responsible, which health category the data falls under, whether the data is
structured, and which legal instrument governs access. Its design is documented by
its authors, who gathered requirements through stakeholder working groups, validated
the result against curated examples in a multi country sandbox, and ran a
pan-European public consultation~\cite{derycke2025}.

That is a design account. It answers whether the profile is fit for purpose. It does
not answer a different and, for anyone planning a national implementation, more
urgent question: how far is the installed base from it right now. This paper answers
that question for the Nordic region, which is the region most often named as
Europe's best case for health data reuse and the region two prominent 2026 documents
propose to build an AI health infrastructure on~\cite{andreassen2026,sitra255}.

\paragraph{Contributions.}
\begin{enumerate}
\item A dated conformance census of the \HdcoCatalogues{} Nordic national catalogues
      harvested by the European data portal, measured against the mandatory
      requirement set that we parse out of the official Release 7 SHACL shapes rather
      than transcribe by hand. Not one of the \HdcoDescriptions{} health themed
      descriptions satisfies all eight mandatory properties, and three of the eight
      are present on exactly zero records.
\item A comparison of that verdict against the portal's own Metadata Quality
      Assessment, which validates DCAT-AP and does not mention the health profile.
      The comparison corrects the reading our own first draft adopted: the health
      layer is not a gap on top of sound generic practice, because the DCAT-AP layer
      beneath is itself only partly conformant, and it also shows that a good MQA
      rating and validated conformance are close to independent.
\item A measurement of theme vocabulary binding across the same catalogues, which
      shows Finland publishing \HdcoFiThemed{} themed descriptions to the European
      portal without a single one bound to the EU theme authority.
\item A defect in the EU data theme authority service. It answers HTTP 200 for any
      IRI in its namespace and returns a well formed but empty RDF document for terms
      it never defined, so \HdcoSquatted{} datasets across the wider portal carry
      theme IRIs that dereference to nothing while passing any status code check.
\item A schema capability analysis of Findata's secondary use catalogue, the richest
      health metadata in the region, showing that six of the eight mandatory
      properties already have a source field and exactly two do not.
\item An OWL~2 vocabulary and three layers of SHACL shapes~\cite{shacl2017} in which
      a conformance verdict is a dated, attributed observation rather than a property
      of the dataset, so that absence, ungrounded values and portal computed values
      are distinguishable. All of it, including the five errors this
      discipline caught, is public~\cite{hdco2026}.
\end{enumerate}

\section{Background}

\subsection{The profile stack}

DCAT~3 is a W3C Recommendation for describing datasets and the catalogues that hold
them~\cite{dcat3}. DCAT-AP is the European application profile over it, constraining
cardinalities and binding several properties to controlled vocabularies published by
the EU Publications Office; version~3.0.1 is current~\cite{dcatap301}. HealthDCAT-AP
is the health extension, developed under the HealthData@EU pilot and published by the
Commission with SHACL shapes for validation~\cite{healthdcatap7}.

Version currency is not a formality here. The specification used to live at
\texttt{healthdcat-ap.github.io}, that page was decommissioned on 22 September 2025,
and its surviving README still names Release~5 as authoritative. Release~6 was
deprecated on 24 April 2026. Release~7 is current and its changelog aligns it with
DCAT-AP~3.0.1. Several third party implementations we encountered still point at the
dead location, so a conformance figure computed today against Release~5 would be
measuring the wrong target. We measure against Release~7.

Release~7 makes eight properties mandatory on \texttt{dcat:Dataset} in both the
public and the restricted layer:

\begin{center}
\begin{tabular}{@{}ll@{}}
\texttt{dct:accessRights} & \texttt{dcatap:applicableLegislation} \\
\texttt{dct:identifier} & \texttt{healthdcatap:hdab} \\
\texttt{dcat:distribution} & \texttt{healthdcatap:healthCategory} \\
\texttt{dcat:theme} & \texttt{healthdcatap:hasStructuredData} \\
\end{tabular}
\end{center}

\noindent
The non-public layer adds \texttt{dcat:contactPoint}, \texttt{dcat:keyword},
\texttt{dct:type} and \texttt{dct:provenance}. \texttt{dct:title} and
\texttt{dct:description} do not appear on that list because the profile inherits
them from DCAT-AP rather than restating them, which is a point worth keeping in
view when reading Figure~\ref{fig:mandatory}.

\subsection{Related work}

Measuring the metadata of open data portals at scale is an established line of
work rather than a new idea, and the honest framing of this paper's contribution
depends on saying exactly where it sits in that line.

Umbrich, Neumaier and Polleres built a public framework for continuously
monitoring metadata quality across Open Data portals~\cite{umbrich2015} and then
generalised it into a schema independent assessment, mapping the metadata of
CKAN, Socrata and OpenDataSoft onto DCAT and reporting findings across more than
260 portals holding 1.1 million datasets~\cite{neumaier2016}. Their metrics are
deliberately portal agnostic: retrievability, completeness against a generic set
of expected fields, accuracy of format declarations, openness of licences. That
generality is the point of the work and it is also its boundary. A metric that
has to hold across three portal software families cannot ask whether a health
dataset declares which health data access body governs it.

Wentzel et al.\ narrowed the target to DCAT-AP specifically and set out the
methodology behind the assessment that data.europa.eu now runs in
production~\cite{wentzel2023}. In the FAIR tradition, F-UJI scores individual
research data objects against operationalised FAIR
metrics~\cite{devaraju2021,wilkinson2016}. Both measure something real and
neither measures profile conformance for a domain extension.

The closest live system is the portal's own Metadata Quality Assessment, which
scores every harvested catalogue on five dimensions with a published points
scheme and four rating bands~\cite{mqa2026}. Its stated scope is compliance with
``DCAT-AP and DCAT-AP derivatives''. HealthDCAT-AP appears nowhere in its
published methodology, and Section~\ref{sec:mqa} reports what happens when its
verdict is placed beside ours on the same eleven catalogues.

Derycke et al.\ report the design and validation of HealthDCAT-AP
itself~\cite{derycke2025}. Their validation ran against curated real world
examples supplied through a multi country sandbox, which is the right test of
whether a profile can express what health catalogues need to say. It is not a
measurement of what live national catalogues publish, and the authors do not
claim it is. National and institutional HealthDCAT-AP implementations exist on
public code hosting, including work from the German Medical Informatics
Initiative; we found none that publishes a conformance census of live
catalogues.

So the gap this paper fills is specific, and it is worth stating narrowly.
Generic portal quality has been measured for a decade. DCAT-AP conformance is
measured continuously and publicly. The health extension that the European
Health Data Space actually depends on has been designed, validated on examples,
and never measured against the installed base. That last measurement is the
contribution here, and everything else in this paper is either the method that
makes it trustworthy or a defect found while performing it.

\subsection{What this work is not}

The Commission publishes an official HealthDCAT-AP validator built on the
Interoperability Test Bed. It checks one record before submission, which is the
right tool for a publisher. We do not reimplement it and this work should not be
used in its place. The shapes reported here do the opposite job: they read a
graph of dated observations covering many catalogues at once and produce a report
addressed to the body that can change the harvest mapping.

\section{Method}

\subsection{Sources}

The primary substrate is the public SPARQL endpoint of the European data portal at
\url{https://data.europa.eu/sparql}, a Virtuoso instance holding 1\,908\,938
\texttt{dcat:Dataset} nodes at the time of the run. The portal harvests national
catalogues, so measuring it measures what the rest of Europe can actually see, which
is the property the Regulation cares about. We cover the \HdcoCatalogues{} Nordic
national catalogues it harvests across Sweden, Denmark, Norway, Finland and Iceland.

The mandatory requirement set is derived from the published shapes rather than typed,
by the procedure given as Algorithm~\ref{alg:reqs}, and a test fails the build if the
committed registry drifts from what that derivation produces.

Findata's Aineistokatalogi, the Finnish secondary use catalogue, is harvested
separately through public read endpoints that the site is not advertising but serves
to every visitor through its client bundle. No access control was circumvented and no
rate limit was evaded. Findata states no reuse licence for this metadata anywhere, so
the harvested payload is not committed; the harvester, the derived aggregate counts
and a small number of individually cited records are.

Two sources failed and the failures constrain what we claim. EUR-Lex answered HTTP
202 with a zero length body on every route tried, including the CELEX HTML and PDF
forms and the ELI form, so the Regulation's implementation dates quoted in
Section~\ref{sec:discussion} rest on the Commission's own summary page and not on the
Official Journal text. Anyone citing chapter or article numbers from this work should
read the primary text first. The Norwegian health register catalogue at
\texttt{helsedata.no} serves no machine readable interface at any path we found, so
Norway is covered through \texttt{data.norge.no} instead.

\subsection{Selection and its bias}

Health themed selection uses \texttt{dcat:theme} equal to the authority term
\texttt{HEAL}. This is a deliberately conservative filter and it has a known
direction of bias: a health dataset that carries no theme, or a theme drawn from a
local vocabulary, is not counted. The \HdcoDescriptions{} figure is therefore a floor
on Nordic health datasets and not an estimate of their number. Finding~2 in
Section~\ref{sec:results} is precisely this effect measured for one country, which is
why we report the two together.

A second consequence is structural. A national health data access body that publishes
no DCAT at all is invisible to this measurement by construction. That is Findata's
position, and rather than treat it as a coverage gap we treat it as the fourth
finding and measure the catalogue directly.

\subsection{Verification}

Nothing in Section~\ref{sec:results} rests on one computation. Three independent
paths run over every headline.

First, dual computation. Every one of \HdcoChecks{} headline figures is computed twice,
once set based over the harvested vectors and once by SPARQL over the emitted graph,
and the run aborts on any disagreement. All \HdcoChecks{} agree in the run reported here. One of them is a
known answer case: the Swedish national portal is checked independently at
\HdcoSwedenKnown{} health descriptions so that a change in the largest contributor
cannot hide inside a regional total.

Second, SHACL, in three gated layers. Layer~1 guards our own recording discipline and
conforms with zero violations. Layer~2 carries one shape per defect class, so the
validation report is the findings table rather than a separate artefact; it reports
\HdcoLayerTwo{} violations, of which \HdcoDefectOne{} are mandatory property absences
and \HdcoDefectTwo{} are health specific absences. Those two classes overlap by
design, because a health specific absence is also a mandatory absence and a reader
usually wants both cuts. \HdcoDefectOne{} is exactly the number of conformance
observations recorded with a false verdict, and \HdcoDefectTwo{} is exactly three
health specific properties across \HdcoDescriptions{} descriptions, so the report
reconciles against the graph rather than merely agreeing with it in tone. Layer~3
holds rules that only bite with several catalogues in the graph at once and reports
\HdcoLayerThree{} violations.

Third, an independent validation engine runs over the OWL core, the generated
requirement registry, the scheme registry and each SHACL layer, and its results are
recorded alongside the pyshacl run.

The emitted graph is \HdcoTriples{} triples. It is written as Turtle text and then
parsed back to prove it is well formed, because building a graph of this size through
an in memory RDF object costs minutes and buys nothing.

\section{Results}
\label{sec:results}

All counts from our own measurement are as at \HdcoDateLong{}. The MQA figures in
Section~\ref{sec:mqa} are a separate snapshot taken on \HdcoMqaRetrieved{}.

\subsection{No Nordic health dataset description is conformant}

\begin{figure}[t]
\centering
\includegraphics[width=\linewidth]{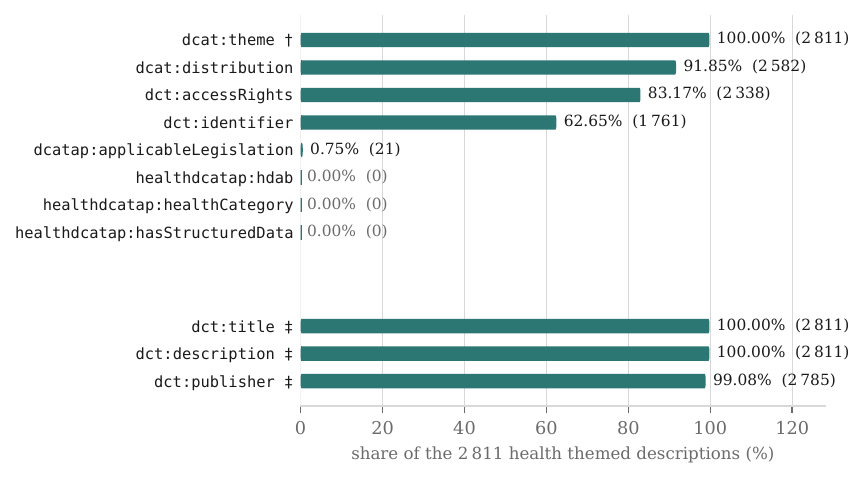}
\caption{Presence of each property across the \HdcoDescriptions{} health themed
dataset descriptions in the \HdcoCatalogues{} Nordic national catalogues.
$\dagger$~\texttt{dcat:theme} is at 100\% by construction of the selection filter
and is not a finding. $\ddagger$~the lower three are inherited from DCAT-AP and
are not restated in the Release 7 mandatory list; they are shown because the
contrast between the two groups is the result.}
\label{fig:mandatory}
\end{figure}

Figure~\ref{fig:mandatory} gives the region wide presence of each mandatory
property. The count of descriptions satisfying all eight is zero, and the useful
result is the shape of the failure rather than the headline.

Three of the eight are present on exactly zero descriptions across five countries
and \HdcoCatalogues{} catalogues, and a fourth,
\texttt{dcatap:applicableLegislation}, reaches \HdcoLegislationPct{}\%. The four
that do carry values are the four DCAT-AP already knew about. Beneath them, the
three properties the profile inherits rather than restates are essentially
complete: title and description on all \HdcoDescriptions{} descriptions,
publisher on \HdcoBasePublisher{}.

The division is not between generic and health specific properties, which is the
reading the figure invites and which we initially adopted. It is between
properties a human types into a publishing form and properties that require a
value drawn from a controlled vocabulary, a legal identifier or an institutional
registration. Title, description and publisher are typed. Health category,
health data access body and applicable legislation must be looked up in something
and bound to it. Section~\ref{sec:mqa} shows that this division cuts through the
DCAT-AP layer too, which is why we do not claim the generic layer is sound.

Zero is a stronger result than a low percentage, because it removes the
possibility that some publishers have solved the problem and others have not.
Nobody in the region has a worked example to copy. That points at a change in
harvest and publication pipelines rather than at a data entry campaign.

\subsection{Finland is absent from European health discovery}

\begin{figure}[t]
\centering
\includegraphics[width=\linewidth]{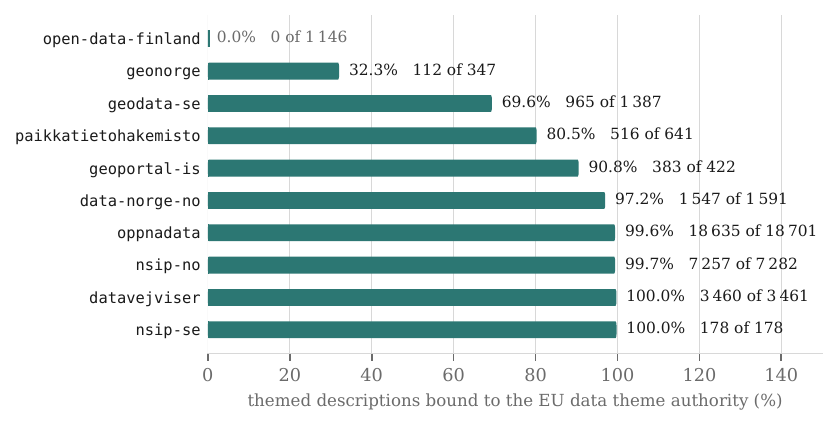}
\caption{Share of themed descriptions whose \texttt{dcat:theme} value is drawn
from the EU data theme authority, by catalogue. \HdcoNoThemeCats{} catalogue of
the \HdcoCatalogues{} publishes no themed description at all and is omitted. A zero is drawn as a tick so that it reads as
measured rather than missing.}
\label{fig:themes}
\end{figure}

Finland's national open data portal contributes \HdcoFiTotal{} dataset
descriptions to the European portal, of which \HdcoFiThemed{} carry a theme. Not
one uses the EU data theme authority vocabulary. Every value is a local CKAN group
identifier under the portal's own domain. Finland is alone in this among the
\HdcoCatalogues{} catalogues, as Figure~\ref{fig:themes} shows: the next lowest
binding rate is \texttt{\HdcoSecondLowestName{}} at \HdcoSecondLowestPct{}\%, and
\HdcoAboveNinety{} of the \HdcoThemedCats{} catalogues that publish any theme at
all are above 90\%.

The operational consequence is direct. A European health filter, which is the
query the Health Data Space is built to serve, returns zero Finnish datasets from
this catalogue. The portal's own search returns 57 results for \emph{health}. The
data is published, indexed at home, and invisible on the route the Regulation
depends on.

This is a one property fix in a harvest mapping and it is the largest gain per
unit of work that the study found. It is also the reason the \HdcoDescriptions{}
figure is a floor: whatever Finnish health datasets exist in that catalogue are
excluded from every other number in this paper.

\subsection{What the portal's own assessment says}
\label{sec:mqa}

The European data portal runs a Metadata Quality Assessment over everything it
harvests, scoring five dimensions out of \HdcoMqaMax{} points and assigning one of
four ratings~\cite{mqa2026}. It validates DCAT-AP. Its published methodology does
not mention HealthDCAT-AP. Placing its verdict beside ours on the same eleven
catalogues (Table~\ref{tab:catalogues}) is therefore not a cross-check of the same
quantity; it is the comparison that locates our result.

\begin{table}[t]
\centering
\small
\caption{Each catalogue as the portal's own DCAT-AP assessment sees it, beside our
HealthDCAT-AP Release 7 verdict. \emph{Datasets} is every dataset the portal holds
for that catalogue and \emph{Health} those carrying the \texttt{HEAL} theme. MQA
scores cover the whole catalogue rather than the health subset and were retrieved
on \HdcoMqaRetrieved{}. \emph{n/r}: the MQA reports no DCAT-AP compliance
indicator for that catalogue.}
\label{tab:catalogues}
\setlength{\tabcolsep}{3.4pt}
\begin{tabular}{llrrrlrr}
\toprule
& & & & \multicolumn{3}{c}{data.europa.eu MQA (DCAT-AP)} & HealthDCAT-AP \\
\cmidrule(lr){5-7}\cmidrule(lr){8-8}
Catalogue & C. & Datasets & Health & Score & Rating & Conf.~(\%) & Conf. \\
\midrule
\texttt{oppnadata} & SE & 23\,340 & 2\,418 & 177 & Sufficient & 27 & 0 \\
\texttt{datavejviser} & DK & 3\,640 & 219 & 287 & Good & 0 & 0 \\
\texttt{data-norge-no} & NO & 1\,637 & 75 & 300 & Good & 62 & 0 \\
\texttt{nsip-no} & NO & 7\,457 & 70 & 248 & Good & 69 & 0 \\
\texttt{geodata-se} & SE & 1\,396 & 19 & 111 & Bad & 0 & 0 \\
\texttt{paikkatietohakemisto} & FI & 641 & 4 & 144 & Sufficient & n/r & 0 \\
\texttt{nsip-se} & SE & 250 & 4 & 191 & Sufficient & 5 & 0 \\
\texttt{geoportal-is} & IS & 483 & 2 & 135 & Sufficient & 0 & 0 \\
\texttt{open-data-finland} & FI & 2\,259 & 0 & 218 & Sufficient & 8 & 0 \\
\texttt{geonorge} & NO & 347 & 0 & 164 & Sufficient & 0 & 0 \\
\texttt{nsip-dk} & DK & 0 & 0 & 252 & Good & 0 & 0 \\
\midrule
Total & & & 2\,811 & & & & 0 \\
\bottomrule
\end{tabular}

\end{table}

Two things follow. First, the DCAT-AP layer underneath is not sound either. The
best Nordic catalogue scores \HdcoMqaBest{} of \HdcoMqaMax{}, none reaches the
\emph{Excellent} band that starts at \HdcoMqaExcellentLow{}, and
\HdcoMqaZeroComp{} of the eleven are reported at zero per cent DCAT-AP compliant.
The largest health contributor by a wide margin, Sweden's national portal, scores
\HdcoMqaSweScore{} with \HdcoMqaSweComp{}\% of its datasets DCAT-AP compliant. So
the honest statement of the headline is not that a health extension has been
skipped on top of solid DCAT-AP practice. It is that the health extension is at
zero on top of a DCAT-AP layer that is itself only partly conformant, and the
health specific properties are the ones with no partial credit available.

Second, and separately useful to anyone reading portal quality scores as
assurance: a good rating and validated conformance are close to independent here.
Denmark's \texttt{datavejviser} scores \HdcoMqaDkScore{} of \HdcoMqaMax{}, a
\emph{Good} rating and the second best in the region, while its DCAT-AP compliance
indicator reads \HdcoMqaDkComp{}\%. The MQA score rewards optional field usage,
resolvable URLs and licence clarity, all of which that catalogue does well. None
of that is conformance. A reader who takes the rating band as evidence that a
catalogue validates will be wrong in the direction that matters.

\subsection{The EU authority namespace accepts terms it never defined}

While measuring theme binding we found a defect one level up, in the EU authority
service itself. Across the whole European portal, well beyond the Nordic scope of the rest
of this paper, \HdcoIrisInUse{} distinct IRIs are in use inside the data theme
namespace. The authority defines \HdcoIrisDefined{} of them, and \HdcoSquatted{}
datasets carry one of the other \HdcoIrisUndefined{}
(Table~\ref{tab:squatting}).

\begin{table}[t]
\centering
\small
\caption{The eight most used undefined terms in the EU data theme namespace, with
the catalogue contributing most of each. All \HdcoIrisUndefined{} undefined terms
account for \HdcoSquatted{} datasets across the whole portal.}
\label{tab:squatting}
\begin{tabular}{lrl}
\toprule
Undefined term & Datasets & Catalogue(s) \\
\midrule
\texttt{undefined} & 806 & \texttt{government-open-data-portal-moldova} (+2) \\
\texttt{ENV} & 237 & \texttt{london-datastore} \\
\texttt{UKLF} & 34 & \texttt{dane-gov-pl} \\
\texttt{VERWALTUNG} & 29 & \texttt{govdata} \\
\texttt{BEVOELKERUNG} & 26 & \texttt{govdata} \\
\texttt{TRANSPORT-VERKEHR} & 23 & \texttt{govdata} \\
\texttt{BILDUNG-WISSENSCHAFT} & 17 & \texttt{govdata} \\
\texttt{KULTUR-FREIZEIT-SPORT-TOURISMUS} & 12 & \texttt{govdata} \\
\midrule
All 22 undefined terms & 1\,238 & \\
\bottomrule
\end{tabular}

\end{table}

The mechanism matters more than the count. The authority host answers HTTP 200 for
any IRI in the namespace and returns a well formed RDF document of
\HdcoEmptyBytes{}~bytes containing no concept for terms it does not define. The
defined term \texttt{HEAL} returns \HdcoHealBytes{}~bytes. A validator that checks
vocabulary membership by dereferencing the IRI and reading the status code
therefore sees a healthy namespace and reports nothing. We nearly shipped that
error ourselves, and membership is now decided by parsing the response body, with
a regression test pinning the behaviour.

Two of the named cases are worth stating plainly because they are cheaply fixable.
The literal string \texttt{undefined} appears as a theme IRI on
\HdcoUndefinedTerm{} datasets, which is a serialisation bug reaching production,
and \HdcoEnvTerm{} more carry \texttt{ENV} where the authority defines
\texttt{ENVI}, a single character transposition in one publisher's mapping.

We also tested whether theme squatting hides a reservoir of health datasets. It
does not. Three datasets across the entire portal carry a German language health
term outside the authority. The general squatting problem is real and large; its
health slice is negligible and we report it as negligible.

\subsection{Findata is the richest catalogue in the region and the least reachable}

Finland's secondary use catalogue holds \HdcoFindataDatasets{} dataset descriptions
and \HdcoFindataVariables{} instance variable descriptions, all with distinct
identifiers. That variable level detail is finer grained than anything in the
Nordic DCAT layer by a wide margin, and it answers the
\texttt{healthdcatap:hasStructuredData} question better than any other catalogue in
the region can.

It is also close to unreachable for a non Finnish speaker.
\HdcoFindataEnLabel{} of the \HdcoFindataDatasets{} descriptions
(\HdcoFindataEnLabelPct{}\%) carry an English label, and
\HdcoFindataNoEnglishVar{} of the \HdcoFindataVariables{} variables
(\HdcoFindataNoEnglishPct{}\%) have no English label at all. The metadata is not
thin. Only \HdcoFindataNoDescVar{} variables (\HdcoFindataNoDescVarPct{}\%) lack a
description. It is substantively complete and monolingual, which for a Regulation
whose stated purpose is cross border discovery is a different failure from
incompleteness and needs a different remedy.

A second reachability defect sits alongside it. All \HdcoFindataConceptTags{}
concept tags in the catalogue report a null concept scheme. That is
\HdcoFindataConcepts{} distinct concepts, \HdcoFindataConceptsEn{} of them English
labelled, bound to no published vocabulary and therefore mappable to nothing.

Table~\ref{tab:findata} asks the question a health data access body would actually
ask: if Findata decided to publish HealthDCAT-AP tomorrow, what could it source
from fields it already has? Six of the eight mandatory properties have a source
field, one of them thin. Exactly two have none. One is
\texttt{dcatap:applicableLegislation}: the catalogue records a usage condition,
which is an access rule, and never the instrument the rule comes from, and while
the Finnish secondary use act appears as free text on some records the string is
not a legal identifier and does not resolve. The other is
\texttt{healthdcatap:hdab}, the health data access body responsible for the
dataset, which for this catalogue is Findata itself. The catalogue cannot state
what it is.

\begin{table}[t]
\centering
\small
\caption{Schema capability of the Findata catalogue against the eight properties
HealthDCAT-AP Release 7 makes mandatory, over \HdcoFindataDatasets{} records.}
\label{tab:findata}
\begin{tabular}{llrrl}
\toprule
Mandatory property & Source field & Records & Share (\%) & Verdict \\
\midrule
\texttt{dct:identifier} & \texttt{id} & 2\,835 & 100.00 & sourceable \\
\texttt{healthdcatap:hasStructuredData} & \texttt{instanceVariables} & 2\,723 & 96.05 & sourceable \\
\texttt{dct:accessRights} & \texttt{usageCondition} & 2\,558 & 90.23 & sourceable \\
\texttt{healthdcatap:healthCategory} & \texttt{datasetTypes} & 2\,542 & 89.66 & sourceable \\
\texttt{dcat:theme} & \texttt{conceptsFromScheme} & 2\,059 & 72.63 & sourceable \\
\texttt{dcat:distribution} & \texttt{links} & 235 & 8.29 & thin \\
\texttt{dcatap:applicableLegislation} & \texttt{--} & 0 & -- & absent \\
\texttt{healthdcatap:hdab} & \texttt{--} & 0 & -- & absent \\
\bottomrule
\end{tabular}

\end{table}

\section{Recording verdicts as observations}

The vocabulary does not describe health data. It describes the descriptions, and the
modelling decision that carries the work is that a conformance verdict is a claim
somebody made, on a date, against a stated requirement, rather than a property of the
dataset.

The load bearing consequence is that absence is recorded, not inferred. A missing
triple and a recorded absence are different things. When a catalogue does not publish
\texttt{healthdcatap:hdab}, that silence is a position the catalogue has taken, and
it is emitted as a conformance observation with a false verdict and a reason drawn
from a controlled scheme, because \emph{absent}, \emph{present but ungrounded} and
\emph{present but not supplied by the publisher} are three different failures with
three different remedies. A model that stores conformance as a boolean on the dataset
cannot tell them apart.

Three observation types carry the results above. A conformance observation binds a
description, a requirement, a verdict and a date. A vocabulary binding observation
records whether a value actually belongs to the scheme the profile declares for it,
which is what makes the Finnish result in Section~\ref{sec:results} expressible at
all: a record can carry every mandatory property and remain undiscoverable because
the values come from a vocabulary no consumer resolves. A language coverage
observation records which languages a property carried, because a description that is
complete domestically and monolingual is not complete for the purpose the Regulation
sets.

A fourth type, provenance observation, exists because of a mistake. See
Section~\ref{sec:errors}.

Requirements are data rather than code. The requirement registry is generated from
the official shapes, and schemes carry their own conformance rules, so the EU data
theme scheme declares its namespace, its closed list flag and its member count and
the pipeline validates against that declaration rather than against a string literal
in a Python file. Changing the profile release changes a generated file, not the
measurement code.

\section{Discussion}
\label{sec:discussion}

\subsection{Two Nordic flagship documents, one missing layer}

Two considered proposals for Nordic health data infrastructure appeared in 2026. The
Sitra study on the future of health data in the age of artificial intelligence
proposes a Finnish Health Data Space on a single infrastructure, a single
authorisation authority and a single research and innovation
body~\cite{sitra255}. The Nature Medicine paper on an AI-Health infrastructure for
the Nordic region proposes a regional platform over linked registries and biobanks, a
Nordic common data model on OMOP, phased federated learning across national trusted
research environments, and integrated high performance computing~\cite{andreassen2026}.
The second cites the first.

Both are explicit that European Health Data Space alignment is a requirement rather
than an afterthought. The Nature Medicine roadmap asks the project to treat EHDS
compliance as a primary design constraint rather than a retrofit, commits the common
data model to rigorous schema auditing and FAIR-compliant
metadata~\cite{wilkinson2016}, and promises schema auditing, ontology mapping and
controlled-access onboarding.

Neither document names DCAT, DCAT-AP or HealthDCAT-AP. Their standardisation is
entirely about what is inside a dataset: OMOP, ICD, imaging, genetics, pathology. The
Sitra text mentions metadata in passing and cites the Finnish authority's own
regulation on dataset descriptions as an instrument, but the discovery profile the
Health Data Space actually runs on does not appear in either.

We state this as an observation about two texts and not as a claim about what their
authors know or have considered. Scoping decisions are legitimate and a content model
paper is not obliged to cover a discovery profile. The observation is still worth
recording, because the layer both documents leave out is the layer where this
measurement returns zero, and a federated platform whose datasets cannot be found
from another member state has solved the harder problem and left the easier one open.

\subsection{What this does not say}

It does not say that any Nordic country is behind on health data. Finland in
particular runs a national health record infrastructure with mandatory deposit from
public and private providers, and a functioning permit process for secondary use.
Nothing measured here contradicts that, and this paper should not be read as a
national scorecard. The measurement covers one axis: the machine readable dataset
catalogue through which another member state discovers that a dataset exists. On that
axis the measured value is zero, and the remedy is a mapping change rather than an
institutional one.

It also does not say that publishers have been careless. Three of the eight
mandatory properties are at zero across five countries and eleven catalogues, which
is what an unimplemented specification looks like rather than what carelessness
looks like. An earlier draft of this paper went further and described the generic
cataloguing as sound, which Section~\ref{sec:mqa} shows is not supportable: the
portal's own assessment puts \HdcoMqaZeroComp{} of these catalogues at zero per cent
DCAT-AP compliance. We record the correction rather than quietly dropping the claim,
because the corrected version carries a different implication for sequencing. If the
DCAT-AP layer is partly conformant, a health profile rollout that assumes a solid
base will fail on the base.

\subsection{Errors we made}
\label{sec:errors}

Five errors were caught before publication and each would have produced a confident,
wrong, published claim. The first four are measurement errors; the fifth is an
over-broad reading of a correct measurement, which is the harder kind to notice. We
report them because the correction procedure is part of the method.

A regular expression over the SHACL file reported a mandatory property set of which
five members are not mandatory. SHACL property shapes are nested blank nodes and a
line oriented regex cannot bracket them. A whole conformance table was computed
against the wrong list before an RDF parse gave the correct set.

A property measured at 98.40\% is in truth at 0.96\%. The triples are present, in
named graphs under the portal's own metrics namespace, which is where the European
data portal stores the output of its own metadata quality assessment. Those triples
were computed by the portal and supplied by no publisher. Every measurement here now
excludes those graphs, and the provenance observation type exists because of it.

A cross check on undefined theme IRIs summed an observed count across every vocabulary
binding observation, which swept in Findata's ungrounded concept tags and reported
4\,186 against a true figure of \HdcoSquatted{}. The dual computation caught it.

HTTP status was nearly used as evidence of vocabulary membership, which as
Section~\ref{sec:results} shows would have reported zero squatted IRIs.

The fifth survived into the first complete draft of this paper and was caught only
when the portal's own assessment was pulled in for comparison. Having measured
title, description and publisher at close to 100\%, we wrote that the generic
cataloguing layer was in good shape. Three inherited properties are not the DCAT-AP
layer, and the portal's own SHACL validation of that layer disagrees sharply.
Section~\ref{sec:mqa} carries the corrected claim. The general lesson is the one
this whole apparatus exists to enforce: a measurement of a subset licenses a
statement about that subset and nothing wider.

\subsection{Threats to validity}

Coverage is limited to catalogues the European data portal harvests, so a national
catalogue publishing no DCAT is invisible by construction. Health themed selection by
authority term biases the count downward by exactly the mechanism reported as the
Finnish result, so \HdcoDescriptions{} is a floor. The Findata harvest returned
\HdcoFindataDatasets{} records against a requested maximum of 5\,000, so it is
complete rather than truncated. Every figure is a single dated observation, which is
why the vocabulary puts a date on every claim: these numbers will move, and the
useful question a year from now is which of them moved.

The MQA figures carry their own limits and are not ours. They were retrieved two days
after the census, on \HdcoMqaRetrieved{}; they score every dataset in a catalogue
rather than the health themed subset; the points scheme is the portal's and rewards
optional field usage alongside validation; and one catalogue reports no DCAT-AP
compliance indicator at all. They are used here to locate our result against the
assessment a reader is most likely to have already seen, not as an independent
confirmation of it.

Finally, the Regulation's implementation dates used as context here could not be read
from the Official Journal text, because EUR-Lex refused every route we tried. They
come from the Commission's own summary page.

\section{Conclusion}

On \HdcoDateLong{}, none of the \HdcoDescriptions{} health themed dataset
descriptions in the \HdcoCatalogues{} Nordic national catalogues harvested by the
European data portal satisfied the requirements HealthDCAT-AP Release~7 places on a
health dataset, and three of the eight mandatory properties were present on exactly
zero records across five countries.

The comparison that gives that number its meaning is with the portal's own quality
assessment. It validates DCAT-AP, it does not mention the health profile, no Nordic
catalogue reaches its top rating band, and \HdcoMqaZeroComp{} of the eleven sit at
zero per cent DCAT-AP compliance. So the result is not a health extension left
undone on an otherwise sound foundation. The properties that survive in both layers
are the ones a publisher types into a form, and the ones that fail in both are the
ones requiring a value bound to a controlled vocabulary, a legal identifier or an
institutional registration. That is a statement about tooling and harvest mappings,
not about diligence, and it means a health profile rollout that assumes a solid
DCAT-AP base will fail on the base.

Two further results are cheaper to fix than the headline and worth more per unit of
effort. Finland publishes \HdcoFiThemed{} themed descriptions to the European portal
with none bound to the EU theme authority, which removes the country from European
health discovery entirely and is a one property change in a harvest mapping. And the
EU authority service itself answers HTTP 200 with an empty document for terms it
never defined, which lets \HdcoSquatted{} datasets carry theme IRIs that resolve to
nothing while passing every status code check a downstream validator is likely to
run.

The vocabulary, the shapes, the harvesters, the generated requirement registry and
the full build report including the five errors described in
Section~\ref{sec:errors} are public~\cite{hdco2026}. The measurement is a single
dated observation by construction, and re-running it is the point.

\section*{Acknowledgements}

This work was carried out independently and received no external funding. We thank
the maintainers of the HealthDCAT-AP specification for publishing the shapes openly,
which is what made an external measurement possible at all.

\section*{Competing interests}

The author is the director of a company that sells consultancy and training in
ontology engineering and data governance, including to public sector buyers in the
domain this paper measures. The measurement code, the vocabulary and the SHACL
shapes are released under an open licence; the derived catalogue mapping artefacts
in the same repository are separately licensed for commercial use. Readers should
weigh the findings against that interest. Every headline in this paper is computed
twice by independent routes and the artefacts are published so the computation can
be repeated without the author.

\section*{Data and code availability}

The vocabulary, the SHACL shapes, the harvesters, the generated requirement registry
and the full build report are public at

\begin{center}
\url{https://github.com/fabio-rovai/health-dataset-catalogue-ontology}
\end{center}

\noindent
Every table and figure in this paper is generated from the committed measurement
artefacts rather than transcribed, so no result number is typed into a section
source and a re-run of the pipeline propagates into the paper unedited. Harvested
payloads are not redistributed: the Findata catalogue metadata carries no stated
reuse licence, and the European portal vectors are regenerable from the public
SPARQL endpoint in about twenty minutes.

\appendix

\section{The measurement, as pseudocode}

The pipeline needs no credentials and runs against three public endpoints: the
European portal's SPARQL service, the authority vocabulary host, and the Findata
catalogue API. Three steps carry the parts where a plausible shortcut gives a wrong
answer, and those are the three given here.

Algorithm~\ref{alg:census} is the census itself. Two details in it are load bearing.
The first is the graph restriction on line~\ref{line:metrics}: the portal stores the
output of its own quality assessment in named graphs under its metrics namespace, and
counting those as publisher supplied moved one property measurement by two orders of
magnitude before we caught it. The second is the commit discipline on
line~\ref{line:commit}. A catalogue enters the cache only when every one of its
property queries has returned, so an interrupted run resumes rather than silently
reporting a partial catalogue as a complete one.

\begin{algorithm}[t]
\caption{Conformance census over a set of catalogues}
\label{alg:census}
\begin{algorithmic}[1]
\Require catalogues $C$, mandatory property set $P$, health theme term $h$
\Ensure presence count $n[c,p]$ for every catalogue and property
\State $\mathit{cache} \gets \emptyset$
\ForAll{$c \in C$}
  \State $G_c \gets \mathit{graphs}(c) \setminus \mathit{metricsGraphs}$ \label{line:metrics}
  \State $D_c \gets \{\, d \in G_c \;:\; d \text{ is a } \texttt{dcat:Dataset} \text{ themed } h \,\}$
  \State $\mathit{complete} \gets \textbf{true}$
  \ForAll{$p \in P$}
    \State $n[c,p] \gets |\{\, d \in D_c \;:\; \exists v.\ (d,p,v) \,\}|$
    \If{the query did not return}
      \State $\mathit{complete} \gets \textbf{false}$
    \EndIf
  \EndFor
  \If{$\mathit{complete}$}
    \State commit $n[c,\cdot]$ to $\mathit{cache}$ \label{line:commit}
  \EndIf
\EndFor
\State \Return $\mathit{cache}$
\end{algorithmic}
\end{algorithm}

Algorithm~\ref{alg:reqs} derives the mandatory property set $P$ that the census
consumes. It is stated separately because transcribing that set by hand, or reading it
off the shapes with a line oriented pattern match, is the error we actually made:
SHACL property constraints hang off the node shape as blank nodes, so a matcher that
cannot bracket a nested node attributes cardinalities to the wrong paths. Parsing the
shapes as RDF is not a stylistic preference here. A regenerated registry is also what
makes a profile release change a data change rather than a code change, and a test
fails the build if the committed registry drifts from what the derivation produces.

\begin{algorithm}[t]
\caption{Deriving the mandatory property set from the published shapes}
\label{alg:reqs}
\begin{algorithmic}[1]
\Require shapes graph $S$ parsed as RDF, target class $t$
\Ensure mandatory property set $P$
\State $P \gets \emptyset$
\ForAll{node shapes $s$ with $(s,\texttt{sh:targetClass},t) \in S$}
  \ForAll{$q$ with $(s,\texttt{sh:property},q) \in S$, where $q$ is a blank node}
    \State $\mathit{path} \gets$ object of $(q,\texttt{sh:path})$
    \State $\mathit{min} \gets$ object of $(q,\texttt{sh:minCount})$, or $0$ if absent
    \If{$\mathit{min} \geq 1$}
      \State $P \gets P \cup \{\mathit{path}\}$
    \EndIf
  \EndFor
\EndFor
\State \Return $P$
\end{algorithmic}
\end{algorithm}

Algorithm~\ref{alg:member} decides whether a theme value belongs to the vocabulary the
profile declares for it. The line that matters is~\ref{line:parse}. Deciding membership
from the HTTP status alone returns true for every IRI in the namespace, including the
\HdcoIrisUndefined{} terms the authority has never defined, because the host answers 200
a well formed but empty document. A regression test pins the body parse so a later
rewrite cannot quietly reintroduce the shortcut.

\begin{algorithm}[t]
\caption{Deciding vocabulary membership for a term}
\label{alg:member}
\begin{algorithmic}[1]
\Require candidate term $t$, authority namespace $N$
\Ensure whether $t$ is defined by the authority
\If{$t \notin N$}
  \State \Return \textbf{false}
\EndIf
\State $r \gets \mathit{dereference}(t)$ with RDF content negotiation
\If{$\mathit{status}(r) \neq 200$}
  \State \Return \textbf{false}
\EndIf
\State $G \gets \mathit{parse}(\mathit{body}(r))$ \label{line:parse}
\State \Return $\exists x.\ (x,\texttt{rdf:type},\texttt{skos:Concept}) \in G$
\end{algorithmic}
\end{algorithm}

Around these three, the remaining steps are routine. The observation graph is emitted
as Turtle text and parsed back to prove it is well formed. Every headline is then
recomputed by both a set based route over the harvested vectors and a SPARQL route
over the emitted graph, and the run aborts on any disagreement. The three shape layers
run last.

The run reported here used Python~3.13, rdflib~7.x and pyshacl, on a graph of
\HdcoTriples{} triples. One caution for anyone re-implementing the cross-source layer:
expressing the claim identity check as a self join over \HdcoObservations{} reified
observations is quadratic and does not terminate in rdflib at this scale. The form we
use compares two counts over a single node target, which is linear and says the same
thing.

\printbibliography

\end{document}